\documentclass[aps,prl,superscriptaddress,twocolumn,nofootinbib,a4paper,amsfonts,amssymb,amsmath,floatfix]{revtex4-2}
\pdfoutput=1
\usepackage[utf8]{inputenc}   
\usepackage[american]{babel}  
\usepackage[
  colorlinks=true,
  allcolors=magenta,
  breaklinks=true
  ]{hyperref}                 
\usepackage[
  caption=false
]{subfig}
\usepackage{bm}		            
\usepackage{microtype}	      
\usepackage{braket}           
\usepackage{tikz}	            
\usetikzlibrary{calc}
\tikzset{event/.style={draw,circle,fill=black,minimum size=8,inner sep=0pt,outer sep=0pt,text=white,font=\tiny}}
\tikzset{nulllike/.style={-,dashed}}
\tikzset{point/.style={fill,draw,circle,minimum size=2,inner sep=0pt,outer sep=0pt}}
\tikzset{vertex/.style={fill,draw,circle,minimum size=3,inner sep=0pt,outer sep=0pt}}
\DeclareMathOperator{\ext}{ext}			 
\DeclareMathOperator{\An}{An}        
\DeclareMathOperator{\Co}{Co}		     
\newcommand{\arc}[0]{\raisebox{0.5ex}{\tikz{\draw[-stealth,thick] (0,0)--(.3,0);}}}
\newcommand{\sarc}[0]{\raisebox{0.4ex}{\tikz{\draw[-stealth] (0,0)--(.2,0);}}}
\newcommand{\edge}[0]{\raisebox{0.5ex}{\tikz{\draw[-,thick] (0,0)--(.3,0);}}}

\begin{document}
\title{The Möbius Game: A Quantum-Inspired Test of General Relativity}
\author{Eleftherios-Ermis Tselentis}
  \affiliation{QuIC, Ecole Polytechnique de Bruxelles, C.P.\ 165, Université Libre de Bruxelles, 1050 Brussels, Belgium}
\author{Ämin Baumeler}
	\affiliation{Facoltà di scienze informatiche, Università della Svizzera italiana, 6900 Lugano, Switzerland}
	\affiliation{Facoltà indipendente di Gandria, 6978 Gandria, Switzerland}

\begin{abstract}
  \noindent
  We present a tight inequality to test {\em the dynamical nature of spacetime.}
  A~general-relativistic violation~of that inequality certifies {\em change of curvature,}
  in the same sense as a~quantum-mechanical~violation of a Bell inequality certifies {\em a source of entanglement.}
  The inequality arises from a minimal generalization of the Bell setup.
  It represents a limit on the winning chance of a collaborative multi-agent game played on {\em the Möbius graph.}
  A long version of this Letter including other games 
  and how these games certify the dynamical character of the celebrated quantum switch 
  is accessible as \href{https://arxiv.org/abs/2309.15752}{arXiv:2309.15752 {\bf [gr-qc].}}
\end{abstract}

\maketitle
Einstein, Podolsky, and Rosen~\cite{einstein1935,bohm1957} challenged the completeness of quantum theory.
For a pair of spin-one-half particles in the singlet state~$(\ket{\uparrow\downarrow}-\ket{\downarrow\uparrow})/\sqrt 2$,
it is possible to {\em perfectly predict\/} any spin component of one of them by {\em interacting with its companion only\/}---even if they are {\em spacelike separated.}
But then, should not the spin components in complementary directions be encoded within each of the particles?
Since the quantum-mechanical description fails to do so, is it possible to complete the theory with hidden variables?
Bell~\cite{bell1964}, most strikingly, answered in the {\em negative:}
The observations in any such theory are severely restricted compared to the quantum-mechanical capabilities.
Suppose we distribute such a ``hidden-variable'' particle to an agent Alice, and another to a spacelike separated agent Bob.
Then, we let each agent perform one out of two dichotomic measurements~$X,Y\in\{0,1\}$ at random, and record the result~$A,B\in\{0,1\}$.
For this scenario, Bell's theorem~\cite{clauser1969} states that their observations are bound by the ``Bell inequality''~$\Pr[A\oplus B=XY]\leq 3/4$ (the symbol~$\oplus$ represents addition modulo two):
The probability for co\"inciding results if and only if at least one agent performed the~$0$ measurement is limited by~$3/4$.
In stark contrast, if the agents were supplied with the quantum particles in the singlet state, then they may {\em violate\/} this inequality and satisfy the predicate with probability~$(2+\sqrt 2)/4$~\cite{tsirelsonsbound}:
These ``nonlocal'' quantum-mechanical correlations are {\em incompatible\/} with local hidden-variable theories;
the quantum-mechanical description of these spin particles cannot be completed.

Bell's discovery is most fascinating not only for its foundational nature, but also for the {\em device-independent methodology:}
Bell's theorem is free of any technological, experimental, or theoretical specifications.
Operations and agents, i.e., experimentalists, are employed as the units of the {\em operational}
 scientific practice drawing the attention into “doings or happenings rather than into objects or entities” \cite{operationalism2}. Any observation that violates a ``Bell inequality'' disagrees with ``local causality''~\cite{bell1990}---{\em the local creation and distribution of information\/} (see, Ref.~\cite{wiseman2017} for a~breakdown of the assumptions needed for Bell's theorem)---irrespective of how the observation came about.
This, in fact, enables applications of highest practical relevance, e.g., the realization of cryptographic tasks without the necessity to trust the device manufacturer~\cite{mayersyao98,bhk05,colbeck2006,scarani12,arnonfriedman19,reviewdi23,cryptoag}.
Note that Bell's theorem may be circumvented, e.g., with hidden variables delocalized in space or time~\cite{bohm1952,costadebeauregard1977,sutherland1983,price1994,baumeler2018a}.
\begin{figure}
  \newcommand{\pastlc}[2]{
    \draw[nulllike] (#1.135) -- ++(180+45:#2);
    \draw[nulllike] (#1.45) -- ++(-45:#2);
  }
  \newcommand{\futurelc}[2]{
    \draw[nulllike] (#1.315) -- ++(45:#2);
    \draw[nulllike] (#1.225) -- ++(135:#2);
  }
  \subfloat[\label{subfig:bell}]{%
    \begin{tikzpicture}
      \node[event] at (0,0) (A) {};
      \node[event] at (1,0) (B) {};
      \node[event,draw=none,fill=none,text=black] at (.5,-.6) (L) {};
      \pastlc{A}{1.5}
      \pastlc{B}{1.5}
    \end{tikzpicture}
  }\hspace{1em}
  \subfloat[\label{subfig:us}]{%
    \begin{tikzpicture}
      \node[event] at (0,0) (A) {};
      \node[event] at (1,0) (B) {};
      \node[event] at (.5,-.6) (L) {};
      \node[event] at (1.25,.5) (C) {};
      \pastlc{A}{1.5}
      \pastlc{B}{1.5}
      \pastlc{C}{1.5}
    \end{tikzpicture}
  }
  \caption{
    (a) In the Bell setup, the agents are spacelike separated. Their observations may depend on some hidden variable generated withing their common past. 
    (b) Here, the agents may be situated in any causal configuration, as long as the causal relations are not tampered with, and the agents may communicate accordingly.
  }
  \label{fig:setup}
\end{figure}
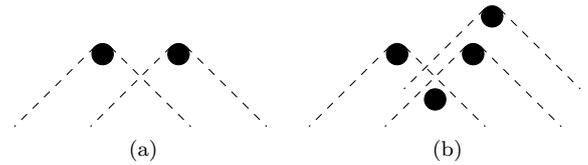

\begin{figure}
  \subfloat[\label{subfig:moebius}]{%
    \includegraphics[scale=.2]{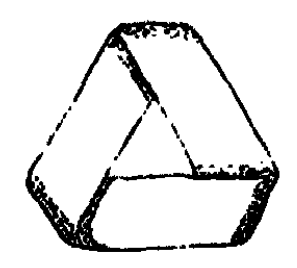}
  }\hspace{4em}
  \subfloat[\label{subfig:moebiusgame}]{%
    \begin{tikzpicture}[every path/.style={-stealth,thick}]
      \def\k{7}
      \def\radius{.7}
      \def\d{.4}
      \pgfmathtruncatemacro{\rot}{90+360/(2*\k)}
      \foreach \i in {1,2,...,\k} {
        \node[vertex] (v\i) at (\rot+360-\i*360/\k:\radius+\d) {};
        \node[vertex] (w\i) at (\rot+360-\i*360/\k:\radius) {};
      }
      \foreach \i in {1,3,...,\k}
      \draw (v\i) -- (w\i);
      \foreach \i in {2,4,...,\k}
      \draw (w\i) -- (v\i);
      \pgfmathtruncatemacro{\km}{\k-1}
      \foreach \i [evaluate=\i as \j using int(\i+1)] in {1,3,...,\km} {
        \draw (w\i) -- (w\j);
        \draw (v\j) -- (v\i);
      }
      \foreach \i [evaluate=\i as \j using int(\i+1)] in {2,4,...,\km} {
        \draw (w\j) -- (w\i);
        \draw (v\i) -- (v\j);
      }
      \draw (w\k) -- (v1);
      \draw (w1) -- (v\k);
    \end{tikzpicture}
  }
  \caption{
    (a) Drawing of a Möbius strip {\em (einseitiger Poly{\"{e}}der)} by August Ferdinand Möbius~\cite[p.~520]{mobius}.
    (b) The 7-Möbius digraph is an orientation of the Möbius ladder~$M_7$.
  }
  \label{fig:moebius}
\end{figure}
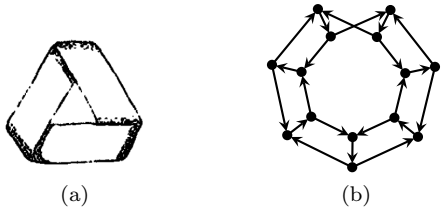

{\it Results.---}In this Letter, we present {\em device-independent tests for the dynamical nature of spacetime,} i.e., whether relativistic correlations are  {\em incompatible with static causal order.}
Here, static causal order means that the causal relations are  unaffected by the agents' actions.
This setup reflects a minimal generalization of Bell's (see Fig.~\ref{fig:setup}).
We describe {\em a relativistic setting within which\/} any violation of the presented inequalities certifies the incompatibility with any static causal background.
In the absence of closed time-like curves~\cite{hawking92}, a violation certifies the dynamical nature of the causal relations; a~general-relativistic violation thus indicates {\em change of curvature.}
This is analogous to the Bell case, where a quantum-mechanical violation of a Bell inequality indicates the presence of entanglement.
The inequalities correspond to a limiting winning chance of collaborative multi-agent games played on directed graphs.
The most promising game discovered is played on an orientation of the Möbius ladder~(see Fig.~\ref{fig:moebius}).
In contrast to other games uncovered in the~longer version of this Letter~\cite{tselentis2023a}, the Möbius game remains non-trivial for any number of agents.
We speculate and argue that such general-relativistic violations of the inequality are feasible.
A concluding assessment, however, requires a~formalization and analysis of the outlined strategy within the framework of general relativity.

{\it The Möbius Game.---}
The game played on the oriented Möbius ladder is fairly simple.
Each vertex on the directed graph represents an agent.
A referee picks at random an arc~$u\arc v$, a~bit~$x$, discloses the selected arc to all agents, and secretly communicates the bit~$x$ to the agent~$u$.
The agents are said {\em to win\/} the game whenever agent~$v$ guesses~$x$.
As we will show, the winning chance under static causal order is upper bounded by 11/12.

{\it Scenario.---}Suppose a referee selects a ``sender''~$s$ and a~distinct ``receiver''~$r$ from the set~\mbox{$[n]:=\{0,1,\dots,n-1\}$} of~$n$ agents, and announces this selection~$(s,r)$ to all agents.
Moreover, the referee provides the ``sender'' with a bit~$x$, and requests from the ``receiver'' a bit~$a$.
Thus, the input space of each agent is the set of distinct agent pairs~$[n]^2\setminus\{(i,i)|i\in[n]\}$, and the ``sender'' has the additional input space~$\{0,1\}$.
The output space of the ``receiver'' is~$\{0,1\}$. 
All other agents have a trivial output space~$\{\bot\}$.
The input-output behavior of this alliance is thus described by the conditional probability distribution~$p(a|s,r,x)$.\footnote{Throughout this Letter we use the same expression for probability distributions and probabilities.}

{\it Static Causal Order.---}Upon reception of the input, each agent {\em immediately\/} responds with the output (which for all-but-one agent is trivial).
We regard the realization of agent~$i$'s output as the {\em event~$E_i$.}
Static causal order is now but a partial order~$\sigma$ over the events~$\{E_i\}_{i\in[n]}$, or, likewise, over~$[n]$.
We use~$i\preceq_\sigma j$ to express that~$E_i$ is in the causal past of~$E_j$ with respect to the partial order~$\sigma$, and~$i \not\preceq_\sigma j$ for the {\em alternative case.}
In the latter case, the output of agent~$j$ {\em cannot\/} depend on the input to agent~$i$.
In the present scenario, the output~$a$---which is produced by the ``receiver''---is independent of the input~$x$---which is provided to the ``sender''---whenever~$s\not\preceq_\sigma r$.
In generality, the partial order~$\sigma$ may be selected probabilistically.
Thus, the input-output behavior~$p(a|s,r,x)$ satisfies {\em static causal order\/} whenever it can be decomposed as a mixture of correlations that satisfy a partial order over the events, i.e., whenever there exists a distribution~$p(\sigma)$ over partial orders, and a family of conditional probability distributions~$\{p_{\sigma}^{\preceq}(a|s,r,x),p_{\sigma}^{\not\preceq}(a|s,r)\}_{\sigma}$, such that
\begin{align}
	\begin{split}
		p(a | s,r,x)
		&=
		\sum_{\sigma: s\preceq_\sigma r} p(\sigma) p_{\sigma}^{\preceq}(a | s,r,x)
		\\+
		&\quad\sum_{\sigma: s\not\preceq_\sigma r} p(\sigma) p_{\sigma}^{\not\preceq}(a | s,r)
		\,.
	\end{split}
	\label{eq:correlations}
\end{align}

{\it Inequality.---}
The Möbius strip is the single-sided surface that is obtained by flipping one end of a~strip and gluing it to the other end (see Fig.~\ref{subfig:moebius}).
The corresponding graph is known as {\em Möbius ladder~$M_k$\/}~\cite{guy1967}, where one end of the~$k$-ladder graph is crosswise connected to the other, i.e.,
it has the vertices~$\{0,1\}\times[k]$, the ``steps''~\mbox{$\{(0,i)\edge(1,i)|i\in[k]\}$}, the ``pole'' edges~$\{(\ell,i)\edge(\ell,i+1)|\ell\in\{0,1\},i\in[k-1]\}$, and the ``crossings''~$\{(0,k-1)\edge(1,0),(1,k-1)\edge(0,0)\}$.
Here, we will consider the {\em directed\/} graph obtained by giving an orientation to each of the edges of the Möbius ladder.
The~{\em $k$-Möbius digraph,} which is defined for odd~$k\geq 3$ only, is~$M_k$ where the edge~$(0,0)\edge(1,0)$ is oriented as~$(0,0)\arc(1,0)$, and where each internal face of the ladder consistently forms a four cycle (see Fig.~\ref{subfig:moebiusgame}).
Also, we refer to any directed graph isomorphic to the~$k$-Möbius digraph---regardless of the vertex names---as~$k$-Möbius digraph.
Note that this digraph has~$3k$ arcs and that it is bipartite, i.e., two colors suffice to paint each of the vertices such that adjacent vertices have different colors.

Now, suppose~$D=(\mathcal V,\mathcal A)$ is a~$k$-Möbius digraph with vertices~$\mathcal V\subseteq[n]$, arcs~$\mathcal A$, and let~$D^{\uparrow n}=([n],\mathcal A)$ be~$D$ where the vertex set is completed to~$[n]$.
Also, let~$\mathcal W(D^{\uparrow n})$ be the event that the~$n$ agents {\em win\/} the M\"obius game when played on~$D^{\uparrow n}$ with uniformly distributed inputs.
Concretely, for some correlations~$p(a|s,r,x)$, the winning chance is
\begin{equation}
  \Pr[\mathcal W(D^{\uparrow n})] = \frac{1}{6k}
  \sum_{\substack{x\in\{0,1\}\\s\sarc r\in\mathcal A}} p(x | s,r,x)
  \,.
  \label{eq:winprob}
\end{equation}
Our main technical contribution is that if the~$n$-agent-correlations~$p(a|s,r,x)$, for any~$n\geq 6$, satisfy static causal order, then this winning chance is bounded by
\begin{equation}
  \Pr[\mathcal W(D^{\uparrow n})]
  \leq 1-\frac{k+1}{12k}
  \,,
  \label{eq:ineq}
\end{equation}
which never exceeds~$11/12$.
This inequality is {\em tight\/} in the strongest sense: It defines a facet of the convex body of correlations~$p(a|s,r,x)$ that satisfy static causal order~(Eq.~\eqref{eq:correlations}).
The above constraint~$n\geq 6$ is due to the fact that the smallest Möbius graph has six vertices.
In the longer version of this Letter, we present further facet-defining inequalities for any number of agents~$n\geq 2$.

{\it Geometric Representation.---}In a first part towards deriving the above inequality, we represent the set of correlations~$p(a|s,r,x)$ that satisfy Eq.~\eqref{eq:correlations} geometrically.
This is done by associating to each such conditional probability distribution the vector~$\bm p:=(p(0|s,r,x))_{(s,r,x)}$.
It is evident that the resulting body~$\mathcal P_n\subseteq\mathbb R^{2n(n-1)}$ forms a~{\em convex polytope,} i.e., it is the convex hull of a finite number of vectors, or, equivalently, it is the intersection of a~finite number of halfspaces.
The inequality~\eqref{eq:ineq} defines such a halfspace.
In the next few paragraphs, we will derive that halfspace from the set~$\ext(\mathcal P_n)$ of extremal vectors of~$\mathcal P_n$.
We achieve this by first {\em projecting\/} the polytope to a lower-dimensional one~$\mathcal Q_n$.
As it turns out, the polytope~$\mathcal Q_n$ is but the polytope of {\em directed acyclic graphs (DAGs).}
Then, we {\em recycle\/} the facet-defining inequalities of~$\mathcal Q_n$, which were derived in Ref.~\cite{acyclic1985}, and lift them to~$\mathcal P_n$.
The reader satisfied with that sketch and not looking for more details may safely skip to the section \hyperlink{par:relativity}{Relativity} where we discuss the relativistic setting.

{\it Properties.---}The polytope~$\mathcal P_n$ has three crucial properties:
It is {\em full-dimensional,\/} all extremal points are {\em deterministic,} and it is what we call {\em ``pairwise centrally symmetric.''}
The first property states that the specification of the~$2n(n-1)$ probabilities are {\em necessary\/} to identify~$p(a|s,r,x)$ within~$\mathcal P_n$.
The second property states~$\ext(\mathcal P_n)\in\{0,1\}^{2n(n-1)}$.
The third property is a~specific form of {\em central symmetry:}
If~$\bm p=(\bm p_0,\bm p_1)$ is an extremal point, 
where we use~$\bm p_b:=(p(0|s,r,b))_{(s,r)}$,
then so is~$(\bm p_0\oplus \bm e_{s',r'}, \bm p_1\oplus \bm e_{s',r'})$ for any~$s',r'$, where~$\bm e_{s',r'}$ is the all-zero vector with a one in 
dimension~$(s',r')$.
To see this, consider the strategy that generates the former correlations, and modify it by letting the ``receiver'' flip the output bit upon input~$s',r'$.

{\it Projection.---}We remove this symmetry of~$\mathcal P_n$ with
\begin{align}
  \pi_n:\{0,1\}^{2n(n-1)} &\rightarrow \{0,1\}^{n(n-1)}\\
  (\bm v_0, \bm v_1) &\mapsto \bm v_0 \oplus \bm v_1
  \,,
\end{align}
and define the resulting polytope~$\mathcal Q_n$ through its extremal points~$\ext(\mathcal Q_n):=\pi_{n}(\ext(\mathcal P_n))$.
The entry in dimension~$(s,r)$ of some extremal~$\bm q=\bm p_0\oplus \bm p_1$ is~$p(0|s,r,0)\oplus p(0|s,r,1)$---it is zero if and only if the output~$a$ of the ``receiver''~$r$ is {\em independent\/} of the input~$x$ to the ``sender''~$s$.
Thus, for each pair~$(s,r)$, the vector~$\bm q$ encodes the dependency of the output of the ``receiver'' from the input to the ``sender.''
Since each such~$\bm q$ arises from a~conditional probability distribution~$p(a|s,r,x)$ satisfying Eq.~\eqref{eq:correlations}, the dependencies must be compatible with a partial order of the agents.
Moreover, since the ``receiver'' in any case may disregard the input to the ``sender,'' the dependencies encoded by~$\bm q$ need not be transitive.
Together this means that~$\bm q$ is the adjacency vector of a {\em DAG.}
Also note that each DAG over the vertices~$[n]$ is represented as an extremal point of~$\mathcal Q_n$.
So, the polytope~$\mathcal Q_n$ is but the polytope of {\em DAGs.}

{\it The DAG Polytope.---}In the context of discrete optimization, Grötschel, Jünger and Reinelt~\cite{acyclic1985} uncovered some of the facet-defining inequalities of~$\mathcal Q_n$.
In particular, they show that for each~$k$-Möbius digraph~\mbox{$D=(\mathcal V\subseteq[n],\mathcal A)$}, the inequality
\begin{equation}
  \sum_{\substack{u\sarc v\in\mathcal A}} q_{u,v}
  \leq
  (5k-1)/2
  \label{eq:gjr}
\end{equation}
is facet-defining for~$\mathcal Q_n$:
If some~\mbox{$\bm q=(q_{u,v})_{(u,v)}\in\mathbb R^{n(n-1)}$} {\em violates\/} that inequality, then~$\bm q$ does not admit a decomposition into a convex combination of the adjacency vectors of DAGs.

{\it Lifting Lemma.---}We present the following lemma, which allows us to recycle the facets of~$\mathcal Q_n$ as facets of~$\mathcal P_n$. If the inequality~$\sum_{i\neq j\in[n]}\chi_{i,j} q_{i,j}\leq c$ is facet-defining for~$\mathcal Q_n$, {\em non-negative,} i.e., all coefficients~$\chi_{(i,j)}$ are non-negative, and {\em non-trivial,} i.e., two or more coefficients are non-zero, then
\begin{equation}
  \sum_{i\neq j\in[n]}\chi_{i,j} p_{i,j,0}
  -
  \sum_{i\neq j\in[n]}\chi_{i,j} p_{i,j,1}
  \leq c
\end{equation}
is a non-trivial facet-defining inequality of~$\mathcal P_n$, where~$p_{i,j,b}$ is the entry in dimension~$(i,j)$ of~$\bm p_b$.

Our main technical contribution---the Möbius inequality~\eqref{eq:ineq}---is obtained by applying that Lifting Lemma (a~proof of which is accessible in the longer version of the present letter~\cite{tselentis2023a}) to the Grötschel-Jünger-Reinelt inequality~\eqref{eq:gjr}.
For a~$k$-Möbius digraph~$D=(\mathcal V\subseteq[n],\mathcal A)$, we thus get
\begin{align}
  \sum_{i\sarc j\in\mathcal A}
  &
  p(0|i,j,0)
  - 
  \sum_{i\sarc j\in\mathcal A}
  p(0|i,j,1)
  \\
  =&
  \sum_{i\sarc j\in\mathcal A}
  \left(
  p(0|i,j,0)
  - 
  (1
  -
  p(1|i,j,1)
  )
  \right)
  \\
  =&
  \sum_{\substack{i\sarc j\in\mathcal A\\b\in\{0,1\}}}
  p(b|i,j,b)
  -
  |\mathcal A|
  \leq
  (5k-1)/2
  \,.
\end{align}
What remains, is to move~$|\mathcal A|$ to the right-hand side, and to multiply the inequality with the uniform distribution of the referee announcing some~$i\arc j$ and~$x$, i.e., with~$(2|\mathcal A|)^{-1}$.

\begin{figure}
  \centering
  \begin{tikzpicture}
    \pgfmathsetmacro{\angle}{45}
    \pgfmathsetmacro{\w}{1}
    \pgfmathsetmacro{\h}{(\w/2)/cos(\angle)}
    \pgfmathsetmacro{\xmax}{3}
    \pgfmathsetmacro{\ymax}{1.5}
    \tikzset{ccc/.style={vertex,fill=white}}
    \foreach \x/\y/\l in {1/.5/L, 2/1/R} {
      \pgfmathsetmacro{\xstart}{\x}
      \pgfmathsetmacro{\ydelta}{tan(\angle)*\w/2}
      \pgfmathsetmacro{\yp}{\y-\ydelta}
      \pgfmathsetmacro{\yf}{\y+\ydelta}
      \coordinate (\l) at (\x,\y);
      \coordinate (P\l) at (\xstart,\yp);
      \coordinate (F\l) at (\xstart,\yf);
      \pgfmathsetmacro{\cleftx}{\xstart/cos(\angle)}
      \pgfmathsetmacro{\clefty}{(\yp-\ymax)/cos(180-\angle)}
      \pgfmathsetmacro{\cleft}{min(\cleftx,\clefty)}
      \pgfmathsetmacro{\crightx}{(\xmax-\xstart)/cos(\angle)}
      \pgfmathsetmacro{\crighty}{(\yp-\ymax)/cos(180-\angle)}
      \pgfmathsetmacro{\cright}{min(\crightx,\crighty)}
      \coordinate (BTL\l) at ($ (\xstart,\yp) + (180-\angle:\cleft) $);
      \coordinate (BTR\l) at ($ (\xstart,\yp) + (\angle:\cright) $);
      \pgfmathsetmacro{\cleftx}{\xstart/sin(\angle)}
      \pgfmathsetmacro{\clefty}{(\yf)/sin(180-\angle)}
      \pgfmathsetmacro{\cleft}{min(\cleftx,\clefty)}
      \pgfmathsetmacro{\crightx}{(\xmax-\xstart)/cos(\angle)}
      \pgfmathsetmacro{\crighty}{\yf/cos(90-\angle)}
      \pgfmathsetmacro{\cright}{min(\crightx,\crighty)}
      \coordinate (BBL\l) at ($ (\xstart,\yf) + (180+\angle:\cleft) $);
      \coordinate (BBR\l) at ($ (\xstart,\yf) + (360-\angle:\cright) $);
    }
    \foreach \l in {L, R} {
      \path[fill=red!20!white] (P\l) -- ++(\angle:\h) -- (F\l) -- ++(180+\angle:\h) -- cycle;
      \node at (\l) {\footnotesize$S_\l$};
    }
    \foreach \l in {L, R} {
      \node[point] at (P\l) {};
      \node[point] at (F\l) {};
      \draw[nulllike] (P\l) -- (BTL\l);
      \draw[nulllike] (P\l) -- (BTR\l);
      \draw[nulllike] (F\l) -- (BBL\l);
      \draw[nulllike] (F\l) -- (BBR\l);
    }
  \end{tikzpicture}
  \caption{Two non-overlapping spacetime regions~$S_L, S_R$ that hold the respective events.}
  \label{fig:regions}
\end{figure}
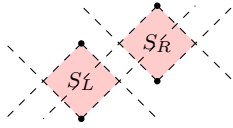

\hypertarget{par:relativity}{{\it Relativity.---}}%
In this section we adopt the Möbius game, as defined via the~$k$-Möbius digraph~$D=([2k],\mathcal A)$, to a~relativistic setting free of any closed time-like curves~\cite{hawking92}.
In the following, a description using a single spatial dimension is sufficient but not necessary.
Violations of the Möbius inequality in this setting certify the dynamical nature of spacetime.\footnote{The same setting may be used for any of the inequalities based on bipartite graphs derived in the longer version of this Letter.}
First, note that if the events~$\{E_i\}_{i\in[2k]}$ are situated at distinct spacetime points in Minkowski spacetime, then it is {\em impossible\/} for the agents to violate the Möbius inequality:
The events form a partial order.
We will now generalize this setting to arrive at a~description suitable for special as well as general relativity, where a special-relativistic violation remains impossible, and where a general-relativistic violation---as we argue below---becomes feasible.
First, it is not necessary for all events to be situated at {\em distinct\/} spacetime points.
Because the~$k$-Möbius digraph is {\em bipartite,} 
we may bipartition the nodes~$[2k]$ into two disjoint sets~$\mathcal V_L,\mathcal V_R$ according to the digraphs' two-coloring.
Events of the same ``color'' may be situated at the same location.
Next, we get rid of absolute spacetime coordinates.
We define two non-overlapping spacetime regions~$S_L$ and $S_R$, which are causal diamonds.
The starting and end points of these diamonds are defined by the intersection of null geodesics.
We confine the events of each ``color'' to the respective region (see Fig.~\ref{fig:regions}).
Any announced arc~$s\arc r$ from the referee will {\em never\/} be such that~$s,r$ are within the same region.
Thus, in a flat but also in a curved spacetime, if the locations of the regions are fixed, then the Möbius inequality cannot be violated.

\begin{figure}
	\centering
  \begin{tikzpicture}
    \tikzset{ccc/.style={vertex,fill=white}}
    \pgfmathsetmacro{\angle}{45}
    \pgfmathsetmacro{\xmax}{3.5}
    \draw[thick] (-\xmax,0) -- (\xmax,0);
    \pgfmathsetmacro{\off}{0.75}
    \pgfmathsetmacro{\xc}{0}
    \pgfmathsetmacro{\xcl}{-\off}
    \pgfmathsetmacro{\xcr}{+\off}
    \pgfmathsetmacro{\xlr}{-2}
    \pgfmathsetmacro{\xll}{\xlr-\off}
    \pgfmathsetmacro{\xrl}{+2}
    \pgfmathsetmacro{\xrr}{\xrl+\off}
    \pgfmathsetmacro{\cmax}{3}
    \foreach \x in {\xlr,\xc,\xrl} {
      \pgfmathsetmacro{\cleft}{min((\xmax+\x)/cos(\angle),\cmax)}
      \pgfmathsetmacro{\cright}{min((\xmax-\x)/cos(\angle),\cmax)}
      \draw[nulllike] (\x,0) -- ++(\angle:\cright);
      \draw[nulllike] (\x,0) -- ++(180-\angle:\cleft);
    }
    \foreach \x in {\xll,\xcl} {
      \pgfmathsetmacro{\cright}{min((\xmax-\x)/cos(\angle),\cmax)}
      \draw[->] (\x,0) -- ++(\angle:\cright);
    }
    \foreach \x in {\xrr,\xcr} {
      \pgfmathsetmacro{\cleft}{min((\xmax+\x)/cos(\angle),\cmax)}
      \draw[->] (\x,0) -- ++(180-\angle:\cleft);
    }
    \pgfmathsetmacro{\xright}{\xrl/2}
    \pgfmathsetmacro{\xleft}{-\xright}
    \pgfmathsetmacro{\xdelta}{\xright}
    \pgfmathsetmacro{\yup}{\xdelta*tan(\angle)}
    \pgfmathsetmacro{\h}{\off/(2*cos(\angle)}
    \pgfmathsetmacro{\yd}{sqrt(2*\h*\h)/2}
    \foreach \x/\y/\l in {\xleft/\yup/L,\xright/\yup/R,0/0/C} {
      \path[fill=red!20!white] (\x,\y) -- ++(\angle:\h) -- ++(180-\angle:\h) -- ++(180+\angle:\h) -- cycle;
      \node at ($ (\x,\y) + (0,\yd) $) {\footnotesize$S_\l$};
    }
    \foreach \x/\l in {\xc/C,\xlr/L,\xrl/R} {
      \node[ccc,label={below:$\An_{\l}$}] (An\l) at (\x,0) {};
    }
    \foreach \x/\l/\s in {\xll/L/\rightarrow,\xrr/R/\rightarrow,\xcl/R/\rightarrow,\xcr/L/\leftarrow} {
      \node[ccc,fill=black,label={below:$\Co^{\s}_{\l}$}] (R\l) at (\x,0) {};
    }
  \end{tikzpicture}
  \caption{Three non-overlapping spacetime regions are defined via intersections of past and future lightcones as constituted by the ``announcers''~$(\An)$ and the ``collectors''~$(\Co)$. The events of the agents~$[2k]=[n]\setminus\{n-1\}$ are partitioned according to their ``color,'' and happen in the regions~$S_L, S_R$ respectively. The event of the extra agent~$n-1$ takes place in the common past within the region~$S_C$.}
	\label{fig:relativity}
\end{figure}
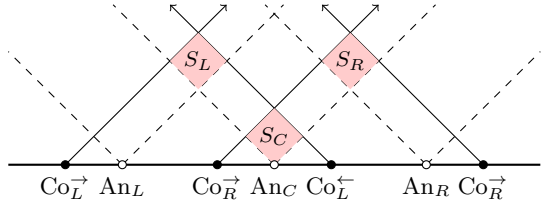

We specify the setting further in order to possibly enable {\em general-relativistic violations\/} of the inequality.
Such violations may occur whenever the causal order between the regions is determined by an event in their common past.
For that purpose, we add one agent without otherwise changing the graph ($n=2k+1$), and locate the event of that additional agent in a spacetime region in the common past of~$S_L$ and~$S_R$.
Also, we may regard~$S_L$ and~$S_R$ as spacelike separated.\footnote{%
The longer version of this Letter~\cite{tselentis2023a} contains an additional setting where the relevant spacetime regions are initially timelike separated.
}
All-in-all, we end up with three spacetime regions in a causal configuration as displayed in Fig.~\ref{fig:relativity}.

Finally, we must ensure that the events take place within the prescribed regions, and that the agents receive their designated input from the referee.
To achieve that, we split the referee into seven referees of two types, ``announcer'' ($\An$) and ``collector'' ($\Co$) who are initially positioned as shown in Fig.~\ref{fig:relativity}.
As their names suggest, the ``announcer'' announce the inputs to the agents, and the ``collectors'' collect the outputs.
The central ``announcer''~$\An_C$ shares a random bit~$b_L$ ($b_R$) with the left (right) ``announcer''~$\An_L$ ($\An_R$), and broadcasts the arc~$s\arc r\in\mathcal A$ together with the bit~$x\oplus b_\chi$, where~$\chi$ is the ``color'' of~$s$.
Both marginal ``announcer'' broadcast the shared bit~$b_L, b_R$.
Therefore, the arc~$s\arc r$ is accessible in the future light cone of the central announcer, i.e., ~$J^+(\An_C)$.
In contrast, if~$s$ has ``color''~$\chi$, then the bit~$x$ is accessible in~$J^+(\An_C)\cap J^+(\An_\chi)$ only.
The ``collectors,'' then again, travel at speed of light in the direction indicated.
All agents of ``color''~$\chi$ must produce the output at the latest when~$\Co^\rightarrow_\chi$ and~$\Co^\leftarrow_\chi$ meet.
Otherwise, the ``collectors'' abort the game.
Our extra agent~$n-1$ must produce the output at the latest when~$\Co^\rightarrow_R$ and~$\Co^\leftarrow_L$ meet.

The rules described above, which are enforced by the referee, ensure that the events~$\{E_i\}_{i\in\mathcal V_\chi}$ happen within the spacetime region
\begin{equation}
  S_\chi:=J^+(\An_C)\cap J^+(\An_\chi)\setminus J^+(\Co^\rightarrow_\chi)\cup J^+(\Co^\leftarrow_\chi)
  \,,
\end{equation}
and~$E_{n-1}$ within~$S_C:=J^+(\An_C)\setminus J^+(Co^\rightarrow_R)\cup J^+(Co^\leftarrow_L)$.
As described above, it is impossible to violate the Möbius inequality while respecting these rules of the game and without dynamically specifying the causal order in every round:
Any theory with a static background must conform to the derived limits.
In contrast, it may be possible to exceed these limits in general relativity for the following reason.
The presence of matter influences the trajectories of particles and light.
Since we define the spacetime regions within which the events occur as causal diamonds, whose starting and end points are defined as the intersection of light beams,
the presence of matter also determines the relative locations of these diamonds, and hence of the events.
Agent~$n-1$, upon reception of the input~$s\arc r$, could in principle specify the distribution of matter within its future light cone, and hence within the common past of the events in~$S_L$ and~$S_R$.
Thus, agent~$n-1$ can specify the causal relation between any pair of events of different ``color.''

Note that general relativity is a deterministic theory, and thus all physical quantities are determined.
The probabilities~$p(a|s, r, x)$ must thus be computed as statistical averages over independent repetitions---being in series or in parallel---under similar boundary conditions.
This is also how Bell experiments are done.
Note that the scenario described is severely more complex when compared to the CHSH-Bell scenario.
This has two reasons: First, we must treat special and general relativity under the same roof, and second, the game outlined is a~``promise'' game, i.e., the inputs~$s,r,x$ cannot be sampled by the agents independently.

{\it Conclusions and Open Questions.---}A statistical violation of the presented inequalities, within the suggested setting, certifies the dynamical nature of spacetime, i.e., the absence of a static causal order.
In relativity, this absence is best described by the backaction of matter on the spacetime manifold.
In fact, a~violation of the presented inequalities within the presented relativistic scenario is a~signature for a~{\em change in curvature}.
A change of curvature, then again, requires the {\em presence and dislocation of matter.}
Thus, a general-relativistic violation of the presented inequalities can also be regarded as a proof of the latter.
One might speculate that the presented approach may also be used to detect {\em gravitational waves.}

Several open questions arise.
Most importantly, while we describe a relativistic setting to violate the Möbius inequality, an explicit description within the general-relativistic framework remains missing.
Such a description may also shed light on the use of the inequality to detect gravitational waves.
Theories with dynamical spacetime have a {\em richer} space of possible correlations. One might therefore wonder about their information-processing capabilities. Can we exploit these correlations to achieve tasks that are impossible in theories with a static background, analogous to the case of quantum information processing~\cite{NielsenChuang2010}?
Does a dynamical spacetime allow for the implementation of cryptographic tasks that are otherwise impossible or not composable~\cite{kent,vilasini}?
On a~technical side, one might wonder whether the Klein bottle, which is a generalization of the Möbius strip, also represents a facet of the polytope of correlations respecting static causal order.

\noindent
{\bf Acknowledgments.}
We thank Luca Apadula, Flavio Del Santo, Andrea Di Biagio, and Stefan Wolf for helpful discussions, and anonymous referees for their valuable comments.
EET thanks Alexandra Elbakyan for providing access to the scientific literature.
EET acknowledges support of the ID\# 61466 and ID\# 62312 grants from the John Templeton Foundation, as part of the ``Quantum Information Structure of Spacetime (QISS)'' project (qiss.fr), and support from the Research Network Quantum Aspects of Spacetime (TURIS).
\"AB is supported by the Swiss National Science Foundation (SNF) through project~214808, and by the Hasler Foundation through project~24010.

\bibliography{references.bib}

\begin{thebibliography}{28}%
\makeatletter
\providecommand \@ifxundefined [1]{%
 \@ifx{#1\undefined}
}%
\providecommand \@ifnum [1]{%
 \ifnum #1\expandafter \@firstoftwo
 \else \expandafter \@secondoftwo
 \fi
}%
\providecommand \@ifx [1]{%
 \ifx #1\expandafter \@firstoftwo
 \else \expandafter \@secondoftwo
 \fi
}%
\providecommand \natexlab [1]{#1}%
\providecommand \enquote  [1]{``#1''}%
\providecommand \bibnamefont  [1]{#1}%
\providecommand \bibfnamefont [1]{#1}%
\providecommand \citenamefont [1]{#1}%
\providecommand \href@noop [0]{\@secondoftwo}%
\providecommand \href [0]{\begingroup \@sanitize@url \@href}%
\providecommand \@href[1]{\@@startlink{#1}\@@href}%
\providecommand \@@href[1]{\endgroup#1\@@endlink}%
\providecommand \@sanitize@url [0]{\catcode `\\12\catcode `\$12\catcode
  `\&12\catcode `\#12\catcode `\^12\catcode `\_12\catcode `\%12\relax}%
\providecommand \@@startlink[1]{}%
\providecommand \@@endlink[0]{}%
\providecommand \url  [0]{\begingroup\@sanitize@url \@url }%
\providecommand \@url [1]{\endgroup\@href {#1}{\urlprefix }}%
\providecommand \urlprefix  [0]{URL }%
\providecommand \Eprint [0]{\href }%
\providecommand \doibase [0]{https://doi.org/}%
\providecommand \selectlanguage [0]{\@gobble}%
\providecommand \bibinfo  [0]{\@secondoftwo}%
\providecommand \bibfield  [0]{\@secondoftwo}%
\providecommand \translation [1]{[#1]}%
\providecommand \BibitemOpen [0]{}%
\providecommand \bibitemStop [0]{}%
\providecommand \bibitemNoStop [0]{.\EOS\space}%
\providecommand \EOS [0]{\spacefactor3000\relax}%
\providecommand \BibitemShut  [1]{\csname bibitem#1\endcsname}%
\let\auto@bib@innerbib\@empty
\bibitem [{\citenamefont {Einstein}\ \emph {et~al.}(1935)\citenamefont
  {Einstein}, \citenamefont {Podolsky},\ and\ \citenamefont
  {Rosen}}]{einstein1935}%
  \BibitemOpen
  \bibfield  {author} {\bibinfo {author} {\bibfnamefont {A.}~\bibnamefont
  {Einstein}}, \bibinfo {author} {\bibfnamefont {B.}~\bibnamefont {Podolsky}},\
  and\ \bibinfo {author} {\bibfnamefont {N.}~\bibnamefont {Rosen}},\ }\bibfield
   {title} {\bibinfo {title} {{Can quantum-mechanical description of physical
  reality be considered complete?}},\ }\href
  {https://doi.org/10.1103/PhysRev.47.777} {\bibfield  {journal} {\bibinfo
  {journal} {Physical Review}\ }\textbf {\bibinfo {volume} {47}},\ \bibinfo
  {pages} {777} (\bibinfo {year} {1935})}\BibitemShut {NoStop}%
\bibitem [{\citenamefont {Bohm}\ and\ \citenamefont
  {Aharonov}(1957)}]{bohm1957}%
  \BibitemOpen
  \bibfield  {author} {\bibinfo {author} {\bibfnamefont {D.}~\bibnamefont
  {Bohm}}\ and\ \bibinfo {author} {\bibfnamefont {Y.}~\bibnamefont
  {Aharonov}},\ }\bibfield  {title} {\bibinfo {title} {{Discussion of
  experimental proof for the paradox of Einstein, Rosen, and Podolsky}},\
  }\href {https://doi.org/10.1103/PhysRev.108.1070} {\bibfield  {journal}
  {\bibinfo  {journal} {Physical Review}\ }\textbf {\bibinfo {volume} {108}},\
  \bibinfo {pages} {1070} (\bibinfo {year} {1957})}\BibitemShut {NoStop}%
\bibitem [{\citenamefont {Bell}(1964)}]{bell1964}%
  \BibitemOpen
  \bibfield  {author} {\bibinfo {author} {\bibfnamefont {J.~S.}\ \bibnamefont
  {Bell}},\ }\bibfield  {title} {\bibinfo {title} {On the {Einstein} {Podolsky}
  {Rosen} paradox},\ }\href
  {https://doi.org/10.1103/PhysicsPhysiqueFizika.1.195} {\bibfield  {journal}
  {\bibinfo  {journal} {Physics Physique Fizika}\ }\textbf {\bibinfo {volume}
  {1}},\ \bibinfo {pages} {195} (\bibinfo {year} {1964})}\BibitemShut {NoStop}%
\bibitem [{\citenamefont {Clauser}\ \emph {et~al.}(1969)\citenamefont
  {Clauser}, \citenamefont {Horne}, \citenamefont {Shimony},\ and\
  \citenamefont {Holt}}]{clauser1969}%
  \BibitemOpen
  \bibfield  {author} {\bibinfo {author} {\bibfnamefont {J.~F.}\ \bibnamefont
  {Clauser}}, \bibinfo {author} {\bibfnamefont {M.~A.}\ \bibnamefont {Horne}},
  \bibinfo {author} {\bibfnamefont {A.}~\bibnamefont {Shimony}},\ and\ \bibinfo
  {author} {\bibfnamefont {R.~A.}\ \bibnamefont {Holt}},\ }\bibfield  {title}
  {\bibinfo {title} {{Proposed experiment to test local hidden-variable
  theories}},\ }\href {https://doi.org/10.1103/PhysRevLett.23.880} {\bibfield
  {journal} {\bibinfo  {journal} {Physical Review Letters}\ }\textbf {\bibinfo
  {volume} {23}},\ \bibinfo {pages} {880} (\bibinfo {year} {1969})}\BibitemShut
  {NoStop}%
\bibitem [{\citenamefont {Cirel'son}(1980)}]{tsirelsonsbound}%
  \BibitemOpen
  \bibfield  {author} {\bibinfo {author} {\bibfnamefont {B.~S.}\ \bibnamefont
  {Cirel'son}},\ }\bibfield  {title} {\bibinfo {title} {{Quantum
  generalizations of Bell's inequality}},\ }\href
  {https://doi.org/10.1007/BF00417500} {\bibfield  {journal} {\bibinfo
  {journal} {Letters in Mathematical Physics}\ }\textbf {\bibinfo {volume}
  {4}},\ \bibinfo {pages} {93} (\bibinfo {year} {1980})}\BibitemShut {NoStop}%
\bibitem [{\citenamefont {Bridgman}(1956)}]{operationalism2}%
  \BibitemOpen
  \bibfield  {author} {\bibinfo {author} {\bibfnamefont {P.~W.}\ \bibnamefont
  {Bridgman}},\ }\bibinfo {title} {Present state of operationalism},\ in\
  \href@noop {} {\emph {\bibinfo {booktitle} {The Validation of Scientific
  Theories}}}\ (\bibinfo  {publisher} {Beacon Press},\ \bibinfo {address}
  {Boston},\ \bibinfo {year} {1956})\ pp.\ \bibinfo {pages}
  {74--79}\BibitemShut {NoStop}%
\bibitem [{\citenamefont {Bell}(1990)}]{bell1990}%
  \BibitemOpen
  \bibfield  {author} {\bibinfo {author} {\bibfnamefont {J.~S.}\ \bibnamefont
  {Bell}},\ }\bibfield  {title} {\bibinfo {title} {{La nouvelle cuisine}},\
  }in\ \href {https://doi.org/10.1016/B978-0-444-88659-0.50010-7} {\emph
  {\bibinfo {booktitle} {Between Science and Technology}}},\ \bibinfo {series
  and number} {North-Holland Delta Series},\ \bibinfo {editor} {edited by\
  \bibinfo {editor} {\bibfnamefont {A.}~\bibnamefont {Sarlemijn}}\ and\
  \bibinfo {editor} {\bibfnamefont {P.}~\bibnamefont {Kroes}}}\ (\bibinfo
  {publisher} {Elsevier},\ \bibinfo {address} {Amsterdam},\ \bibinfo {year}
  {1990})\ Chap.~\bibinfo {chapter} {6}, pp.\ \bibinfo {pages}
  {97--115}\BibitemShut {NoStop}%
\bibitem [{\citenamefont {Wiseman}\ and\ \citenamefont
  {Cavalcanti}(2017)}]{wiseman2017}%
  \BibitemOpen
  \bibfield  {author} {\bibinfo {author} {\bibfnamefont {H.~M.}\ \bibnamefont
  {Wiseman}}\ and\ \bibinfo {author} {\bibfnamefont {E.~G.}\ \bibnamefont
  {Cavalcanti}},\ }\bibfield  {title} {\bibinfo {title} {{Causarum investigatio
  and the two Bell's theorems of John Bell}},\ }in\ \href
  {https://doi.org/10.1007/978-3-319-38987-5\_6} {\emph {\bibinfo {booktitle}
  {Quantum [Un]Speakables II}}},\ \bibinfo {series and number} {The Frontiers
  Collection},\ \bibinfo {editor} {edited by\ \bibinfo {editor} {\bibfnamefont
  {R.}~\bibnamefont {Bertlmann}}\ and\ \bibinfo {editor} {\bibfnamefont
  {A.}~\bibnamefont {Zeilinger}}}\ (\bibinfo  {publisher} {Springer
  International Publishing},\ \bibinfo {address} {Cham},\ \bibinfo {year}
  {2017})\ Chap.~\bibinfo {chapter} {6}, pp.\ \bibinfo {pages}
  {119--142}\BibitemShut {NoStop}%
\bibitem [{\citenamefont {Mayers}\ and\ \citenamefont
  {Yao}(1998)}]{mayersyao98}%
  \BibitemOpen
  \bibfield  {author} {\bibinfo {author} {\bibfnamefont {D.}~\bibnamefont
  {Mayers}}\ and\ \bibinfo {author} {\bibfnamefont {A.}~\bibnamefont {Yao}},\
  }\bibfield  {title} {\bibinfo {title} {Quantum cryptography with imperfect
  apparatus},\ }in\ \href {https://doi.org/10.1109/sfcs.1998.743501} {\emph
  {\bibinfo {booktitle} {39th Annual Symposium on Foundations of Computer
  Science}}}\ (\bibinfo  {publisher} {IEEE},\ \bibinfo {year} {1998})\ pp.\
  \bibinfo {pages} {503--509}\BibitemShut {NoStop}%
\bibitem [{\citenamefont {Barrett}\ \emph {et~al.}(2005)\citenamefont
  {Barrett}, \citenamefont {Hardy},\ and\ \citenamefont {Kent}}]{bhk05}%
  \BibitemOpen
  \bibfield  {author} {\bibinfo {author} {\bibfnamefont {J.}~\bibnamefont
  {Barrett}}, \bibinfo {author} {\bibfnamefont {L.}~\bibnamefont {Hardy}},\
  and\ \bibinfo {author} {\bibfnamefont {A.}~\bibnamefont {Kent}},\ }\bibfield
  {title} {\bibinfo {title} {No signaling and quantum key distribution},\
  }\href {https://doi.org/10.1103/physrevlett.95.010503} {\bibfield  {journal}
  {\bibinfo  {journal} {Physical Review Letters}\ }\textbf {\bibinfo {volume}
  {95}},\ \bibinfo {pages} {010503} (\bibinfo {year} {2005})}\BibitemShut
  {NoStop}%
\bibitem [{\citenamefont {Colbeck}(2006)}]{colbeck2006}%
  \BibitemOpen
  \bibfield  {author} {\bibinfo {author} {\bibfnamefont {R.}~\bibnamefont
  {Colbeck}},\ }\emph {\bibinfo {title} {{Quantum and relativistic protocols
  for secure multi-party computation}}},\ \href
  {http://arxiv.org/abs/0911.3814} {Ph.D. thesis},\ \bibinfo  {school}
  {University of Cambridge} (\bibinfo {year} {2006}),\ \Eprint
  {https://arxiv.org/abs/0911.3814} {arXiv:0911.3814} \BibitemShut {NoStop}%
\bibitem [{\citenamefont {Scarani}(2012)}]{scarani12}%
  \BibitemOpen
  \bibfield  {author} {\bibinfo {author} {\bibfnamefont {V.}~\bibnamefont
  {Scarani}},\ }\bibfield  {title} {\bibinfo {title} {The device-independent
  outlook on quantum physics},\ }\href
  {https://doi.org/10.2478/v10155-012-0003-4} {\bibfield  {journal} {\bibinfo
  {journal} {acta physica slovaca}\ }\textbf {\bibinfo {volume} {62}},\
  \bibinfo {pages} {347} (\bibinfo {year} {2012})}\BibitemShut {NoStop}%
\bibitem [{\citenamefont {Arnon-Friedman}\ \emph {et~al.}(2019)\citenamefont
  {Arnon-Friedman}, \citenamefont {Renner},\ and\ \citenamefont
  {Vidick}}]{arnonfriedman19}%
  \BibitemOpen
  \bibfield  {author} {\bibinfo {author} {\bibfnamefont {R.}~\bibnamefont
  {Arnon-Friedman}}, \bibinfo {author} {\bibfnamefont {R.}~\bibnamefont
  {Renner}},\ and\ \bibinfo {author} {\bibfnamefont {T.}~\bibnamefont
  {Vidick}},\ }\bibfield  {title} {\bibinfo {title} {Simple and tight
  device-independent security proofs},\ }\href
  {https://doi.org/10.1137/18m1174726} {\bibfield  {journal} {\bibinfo
  {journal} {SIAM Journal on Computing}\ }\textbf {\bibinfo {volume} {48}},\
  \bibinfo {pages} {181} (\bibinfo {year} {2019})}\BibitemShut {NoStop}%
\bibitem [{\citenamefont {Zapatero}\ \emph {et~al.}(2023)\citenamefont
  {Zapatero}, \citenamefont {van Leent}, \citenamefont {Arnon-Friedman},
  \citenamefont {Liu}, \citenamefont {Zhang}, \citenamefont {Weinfurter},\ and\
  \citenamefont {Curty}}]{reviewdi23}%
  \BibitemOpen
  \bibfield  {author} {\bibinfo {author} {\bibfnamefont {V.}~\bibnamefont
  {Zapatero}}, \bibinfo {author} {\bibfnamefont {T.}~\bibnamefont {van Leent}},
  \bibinfo {author} {\bibfnamefont {R.}~\bibnamefont {Arnon-Friedman}},
  \bibinfo {author} {\bibfnamefont {W.-Z.}\ \bibnamefont {Liu}}, \bibinfo
  {author} {\bibfnamefont {Q.}~\bibnamefont {Zhang}}, \bibinfo {author}
  {\bibfnamefont {H.}~\bibnamefont {Weinfurter}},\ and\ \bibinfo {author}
  {\bibfnamefont {M.}~\bibnamefont {Curty}},\ }\bibfield  {title} {\bibinfo
  {title} {Advances in device-independent quantum key distribution},\ }\href
  {https://doi.org/10.1038/s41534-023-00684-x} {\bibfield  {journal} {\bibinfo
  {journal} {npj Quantum Information}\ }\textbf {\bibinfo {volume} {9}},\
  \bibinfo {pages} {10} (\bibinfo {year} {2023})}\BibitemShut {NoStop}%
\bibitem [{\citenamefont {Miller}(2020)}]{cryptoag}%
  \BibitemOpen
  \bibfield  {author} {\bibinfo {author} {\bibfnamefont {G.}~\bibnamefont
  {Miller}},\ }\href
  {https://www.washingtonpost.com/graphics/2020/world/national-security/cia-crypto-encryption-machines-espionage/}
  {\bibinfo {title} {{The intelligence coup of the century: How the CIA used
  Crypto AG encryption devices to spy on countries for decades}}},\ \bibinfo
  {howpublished} {The Washington Post} (\bibinfo {year} {2020})\BibitemShut
  {NoStop}%
\bibitem [{\citenamefont {Bohm}(1952)}]{bohm1952}%
  \BibitemOpen
  \bibfield  {author} {\bibinfo {author} {\bibfnamefont {D.}~\bibnamefont
  {Bohm}},\ }\bibfield  {title} {\bibinfo {title} {{A suggested interpretation
  of the quantum theory in terms of ``hidden'' variables. I}},\ }\href
  {https://doi.org/10.1103/PhysRev.85.166} {\bibfield  {journal} {\bibinfo
  {journal} {Physical Review}\ }\textbf {\bibinfo {volume} {85}},\ \bibinfo
  {pages} {166} (\bibinfo {year} {1952})}\BibitemShut {NoStop}%
\bibitem [{\citenamefont {Costa~de Beauregard}(1977)}]{costadebeauregard1977}%
  \BibitemOpen
  \bibfield  {author} {\bibinfo {author} {\bibfnamefont {O.}~\bibnamefont
  {Costa~de Beauregard}},\ }\bibfield  {title} {\bibinfo {title} {{Time
  symmetry and the Einstein paradox}},\ }\href
  {https://doi.org/10.1007/BF02906749} {\bibfield  {journal} {\bibinfo
  {journal} {Il Nuovo Cimento B}\ }\textbf {\bibinfo {volume} {42}},\ \bibinfo
  {pages} {41} (\bibinfo {year} {1977})}\BibitemShut {NoStop}%
\bibitem [{\citenamefont {Sutherland}(1983)}]{sutherland1983}%
  \BibitemOpen
  \bibfield  {author} {\bibinfo {author} {\bibfnamefont {R.~I.}\ \bibnamefont
  {Sutherland}},\ }\bibfield  {title} {\bibinfo {title} {{Bell's theorem and
  backwards-in-time causality}},\ }\href {https://doi.org/10.1007/BF02082904}
  {\bibfield  {journal} {\bibinfo  {journal} {International Journal of
  Theoretical Physics}\ }\textbf {\bibinfo {volume} {22}},\ \bibinfo {pages}
  {377} (\bibinfo {year} {1983})}\BibitemShut {NoStop}%
\bibitem [{\citenamefont {Price}(1994)}]{price1994}%
  \BibitemOpen
  \bibfield  {author} {\bibinfo {author} {\bibfnamefont {H.}~\bibnamefont
  {Price}},\ }\bibfield  {title} {\bibinfo {title} {{A neglected route to
  realism about quantum mechanics}},\ }\href
  {https://doi.org/10.1093/mind/103.411.303} {\bibfield  {journal} {\bibinfo
  {journal} {Mind}\ }\textbf {\bibinfo {volume} {103}},\ \bibinfo {pages} {303}
  (\bibinfo {year} {1994})}\BibitemShut {NoStop}%
\bibitem [{\citenamefont {Baumeler}\ \emph {et~al.}(2018)\citenamefont
  {Baumeler}, \citenamefont {Degorre},\ and\ \citenamefont
  {Wolf}}]{baumeler2018a}%
  \BibitemOpen
  \bibfield  {author} {\bibinfo {author} {\bibfnamefont {{\"{A}}.}~\bibnamefont
  {Baumeler}}, \bibinfo {author} {\bibfnamefont {J.}~\bibnamefont {Degorre}},\
  and\ \bibinfo {author} {\bibfnamefont {S.}~\bibnamefont {Wolf}},\ }\bibfield
  {title} {\bibinfo {title} {{Bell correlations and the common future}},\ }in\
  \href {https://doi.org/10.1007/978-3-319-74971-6_18} {\emph {\bibinfo
  {booktitle} {Quantum Foundations, Probability and Information}}},\ \bibinfo
  {editor} {edited by\ \bibinfo {editor} {\bibfnamefont {A.}~\bibnamefont
  {Khrennikov}}\ and\ \bibinfo {editor} {\bibfnamefont {B.}~\bibnamefont
  {Toni}}}\ (\bibinfo  {publisher} {Springer International Publishing},\
  \bibinfo {address} {Cham},\ \bibinfo {year} {2018})\ pp.\ \bibinfo {pages}
  {255--268}\BibitemShut {NoStop}%
\bibitem [{\citenamefont {Hawking}(1992)}]{hawking92}%
  \BibitemOpen
  \bibfield  {author} {\bibinfo {author} {\bibfnamefont {S.~W.}\ \bibnamefont
  {Hawking}},\ }\bibfield  {title} {\bibinfo {title} {Chronology protection
  conjecture},\ }\href {https://doi.org/10.1103/physrevd.46.603} {\bibfield
  {journal} {\bibinfo  {journal} {Physical Review D}\ }\textbf {\bibinfo
  {volume} {46}},\ \bibinfo {pages} {603} (\bibinfo {year} {1992})}\BibitemShut
  {NoStop}%
\bibitem [{\citenamefont {Tselentis}\ and\ \citenamefont
  {Baumeler}(2023)}]{tselentis2023a}%
  \BibitemOpen
  \bibfield  {author} {\bibinfo {author} {\bibfnamefont {E.-E.}\ \bibnamefont
  {Tselentis}}\ and\ \bibinfo {author} {\bibfnamefont {{\"{A}}.}~\bibnamefont
  {Baumeler}},\ }\bibfield  {title} {\bibinfo {title} {{The M{\"{o}}bius game
  and other Bell tests for relativity}},\ }\Eprint
  {https://arxiv.org/abs/2309.15752} {arXiv:2309.15752}  (\bibinfo {year}
  {2023}),\ \bibinfo {note} {preprint}\BibitemShut {NoStop}%
\bibitem [{\citenamefont {M{\"{o}}bius}(1886)}]{mobius}%
  \BibitemOpen
  \bibfield  {author} {\bibinfo {author} {\bibfnamefont {A.~F.}\ \bibnamefont
  {M{\"{o}}bius}},\ }\href
  {http://sites.mathdoc.fr/cgi-bin/oetoc?id=OE{\_}MOBIUS{\_}{\_}2} {\emph
  {\bibinfo {title} {{Gesammelte Werke: Zweiter Band}}}},\ edited by\ \bibinfo
  {editor} {\bibfnamefont {F.}~\bibnamefont {Klein}}\ (\bibinfo  {publisher}
  {S.~Hirzel},\ \bibinfo {address} {Leipzig},\ \bibinfo {year}
  {1886})\BibitemShut {NoStop}%
\bibitem [{\citenamefont {Guy}\ and\ \citenamefont {Harary}(1967)}]{guy1967}%
  \BibitemOpen
  \bibfield  {author} {\bibinfo {author} {\bibfnamefont {R.~K.}\ \bibnamefont
  {Guy}}\ and\ \bibinfo {author} {\bibfnamefont {F.}~\bibnamefont {Harary}},\
  }\bibfield  {title} {\bibinfo {title} {On the {M{\"{o}}bius} ladders},\
  }\href {https://doi.org/10.4153/cmb-1967-046-4} {\bibfield  {journal}
  {\bibinfo  {journal} {Canadian Mathematical Bulletin}\ }\textbf {\bibinfo
  {volume} {10}},\ \bibinfo {pages} {493} (\bibinfo {year} {1967})}\BibitemShut
  {NoStop}%
\bibitem [{\citenamefont {Gr{\"o}tschel}\ \emph {et~al.}(1985)\citenamefont
  {Gr{\"o}tschel}, \citenamefont {J{\"u}nger},\ and\ \citenamefont
  {Reinelt}}]{acyclic1985}%
  \BibitemOpen
  \bibfield  {author} {\bibinfo {author} {\bibfnamefont {M.}~\bibnamefont
  {Gr{\"o}tschel}}, \bibinfo {author} {\bibfnamefont {M.}~\bibnamefont
  {J{\"u}nger}},\ and\ \bibinfo {author} {\bibfnamefont {G.}~\bibnamefont
  {Reinelt}},\ }\bibfield  {title} {\bibinfo {title} {On the acyclic subgraph
  polytope},\ }\href {https://doi.org/10.1007/BF01582009} {\bibfield  {journal}
  {\bibinfo  {journal} {Mathematical Programming}\ }\textbf {\bibinfo {volume}
  {33}},\ \bibinfo {pages} {28} (\bibinfo {year} {1985})}\BibitemShut {NoStop}%
\bibitem [{\citenamefont {Nielsen}\ and\ \citenamefont
  {Chuang}(2010)}]{NielsenChuang2010}%
  \BibitemOpen
  \bibfield  {author} {\bibinfo {author} {\bibfnamefont {M.~A.}\ \bibnamefont
  {Nielsen}}\ and\ \bibinfo {author} {\bibfnamefont {I.~L.}\ \bibnamefont
  {Chuang}},\ }\href {https://doi.org/10.1017/CBO9780511976667} {\emph
  {\bibinfo {title} {Quantum Computation and Quantum Information: 10th
  Anniversary Edition}}}\ (\bibinfo  {publisher} {Cambridge University Press},\
  \bibinfo {address} {Cambridge},\ \bibinfo {year} {2010})\BibitemShut
  {NoStop}%
\bibitem [{\citenamefont {Kent}(1999)}]{kent}%
  \BibitemOpen
  \bibfield  {author} {\bibinfo {author} {\bibfnamefont {A.}~\bibnamefont
  {Kent}},\ }\bibfield  {title} {\bibinfo {title} {Unconditionally secure bit
  commitment},\ }\href {https://doi.org/10.1103/PhysRevLett.83.1447} {\bibfield
   {journal} {\bibinfo  {journal} {Physical Review Letters}\ }\textbf {\bibinfo
  {volume} {83}},\ \bibinfo {pages} {1447} (\bibinfo {year}
  {1999})}\BibitemShut {NoStop}%
\bibitem [{\citenamefont {Vilasini}\ \emph {et~al.}(2019)\citenamefont
  {Vilasini}, \citenamefont {Portmann},\ and\ \citenamefont {del
  Rio}}]{vilasini}%
  \BibitemOpen
  \bibfield  {author} {\bibinfo {author} {\bibfnamefont {V.}~\bibnamefont
  {Vilasini}}, \bibinfo {author} {\bibfnamefont {C.}~\bibnamefont {Portmann}},\
  and\ \bibinfo {author} {\bibfnamefont {L.}~\bibnamefont {del Rio}},\
  }\bibfield  {title} {\bibinfo {title} {Composable security in relativistic
  quantum cryptography},\ }\href {https://doi.org/10.1088/1367-2630/ab0e3b}
  {\bibfield  {journal} {\bibinfo  {journal} {New Journal of Physics}\ }\textbf
  {\bibinfo {volume} {21}},\ \bibinfo {pages} {043057} (\bibinfo {year}
  {2019})}\BibitemShut {NoStop}%
\end{thebibliography}%
\end{document}